# Observation of Acoustic Non-Hermitian Bloch Braids and Associated Topological Phase Transitions

Qicheng Zhang, Yitong Li, Huanfa Sun, Xun Liu, Luekai Zhao, Xiling Feng, Xiying Fan, and Chunyin Qiu*

Key Laboratory of Artificial Micro- and Nano-Structures of Ministry of Education and School of Physics and Technology

Wuhan University, Wuhan 430072, China

* To whom correspondence should be addressed: cyqiu@whu.edu.cn

*Abstract*. Topological features embedded in ancient braiding and knotting arts endow significant impacts on our daily life and even cutting-edge science. Recently, fast growing efforts are invested to the braiding topology of complex Bloch bands in non-Hermitian systems. This new classification of band topology goes far beyond those established in Hermitian counterparts. Here, we present the first acoustic realization of the topological non-Hermitian Bloch braids, based on a two-band model easily accessible for realizing any desired knot structure. The non-Hermitian bands are synthesized by a simple binary cavity-tube system, where the long-range, complex-valued, and momentum-resolved couplings are accomplished by a well-controlled unidirectional coupler. In addition to directly visualizing various two-band braiding patterns, we unambiguously observe the highly-elusive topological phase transitions between them. Not only do our results provide a direct demonstration for the non-Hermitian band topology, but also the experimental techniques open new avenues for designing unconventional acoustic metamaterials.

*Introduction.* Braids and knots are ubiquitous in nature [1], from the simple shoelace knots to the mysterious DNA molecular knots [2]. All knots can be formed by twisting the strings and joining their ends together, however, they are classified by distinct braiding topology. Figure 1 exemplifies the simplest one-dimensional (1D) topological braids formed by double strands. As sketched in each dashed circle, the two strands inside the cylinders vary from unbraiding to braiding $v$ times, which form respectively an unlink ($v = 0$), an unknot ($v = 1$), a Hopf link ($v = 2$), and a Trefoil knot ($v = 3$) by joining each cylinder into a torus. Obviously, these links/knots belong to different topological structures because they cannot be continuously deformed into each other, unless experiencing a transition state that features one or multiple touching points between the strands.

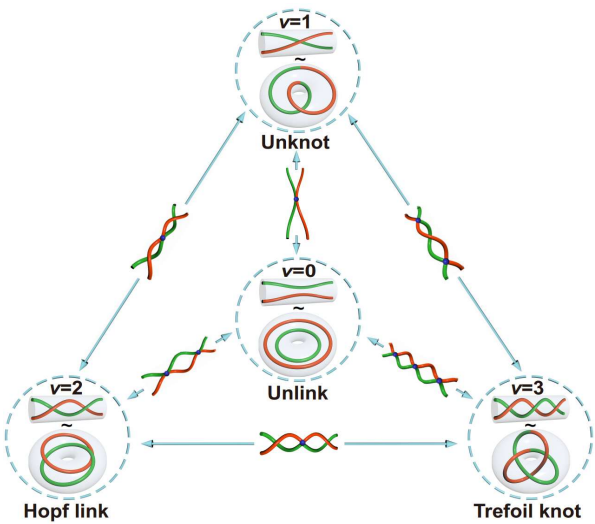

FIG. 1. Double-strand braid patterns and the transition states in between. As sketched in each dashed circle, the two strands inside the cylinder form a link or knot structure when closing the braid into a torus. Topological distinct braids can transit each other through an intermediate state with isolated touching points (blue spheres). The picture is preserved in 1D NH Bloch band systems, where each band can be regarded as a strand of a braid in the complex *E-k* space.

Physical systems, from condensed matter physics [3-8] to photonics [9-13] and acoustics [14,15], also exhibit rich topological braiding structures. For the widely studied Hermitian systems, in addition to the most intuitive multimode braiding in real space [16,17], band node entanglements are unveiled in two- and three-dimensional topological semimetals [5,6,14,15,18,19]. The latter is of particular interest since it reflects intricate Hermitian band topology in high-dimensional energy-momentum (*E-k*) space. By contrast, non-Hermitian (NH) Bloch bands can carry abundant winding and braiding topologies in the complex *E-k* space, even in 1D systems [8,12,20-28]. Specifically, two 1D NH bands can be analogized as the double strands inside the cylinder, where the *E*-plane and *k* are mapped to the cylindrical cross section and axis in Fig. 1, respectively. Very recently, theoretical studies have further revealed that the NH topology of separable bands in a lattice system without additional symmetry can be classified by the conjugacy class of the braid group $\mathbb{B}_N$ [23-25,29,30], enabling a new topological classification beyond that based on line or point gaps [20-22].

Note that much different from the widely studied NH physics in parameter space, which focuses on the unique properties of exceptional points (EPs) [31-41], here the NH band topology considers the evolution of the eigenvalues/eigenstates in momentum space. Interestingly, the control space (i.e., torus-like Brillouin zone) is topologically *nontrivial* in its nature, in contrast to the previous (usually) *trivial* parameter space [13]. Nonetheless, so far only few experiments have been reported on the braiding topology of NH Bloch bands [8,12], comparing to the prosperous studies of the NH physics in parameter space. In particular, the intermediate phases between topologically distinct Bloch braids, featuring unique band degeneracies (i.e., EPs) in the complex band structures, have been long-desired yet not demonstrated in any experiment.

Here, we report the first acoustic realization of the NH two-band topological braids and their phase transitions in between. We start with a simple two-band model that is friendly to the acoustic realization of an arbitrary braiding degree [25]. Experimentally, we employ the concept of synthetic dimensions [12,27,42] to demonstrate the highly-intricate braiding topology inherited in 1D NH Bloch bands, instead of using a finite system where the truncated boundary drastically modifies the energy spectrum and eigenstates of the corresponding infinite lattice [43-46]. We consider a binary-cavity structure equipped with a well-controlled unidirectional coupler. The latter, consisting of an external amplifier and a phase modulator, is elaborately designed to achieve the long-range, complex-valued, and momentum-resolved couplings that are still extremely challenging in acoustics. (Note that although multiple techniques have been developed to achieve acoustic nonreciprocity, e.g., by utilizing the flow-induced bias and active feedback controls [28,47,48], a nonreciprocal coupling with simultaneous amplitude and phase modulations has not yet been reported in acoustics.) Eventually, the braiding physics is unveiled through probing acoustic NH bands in synthetic space, where the lattice momentum is synthesized by the flexibly tunable nonreciprocal coupling. From both the perspectives of eigenvalues and eigenstates, we have not only observed a series of nontrivial braid patterns formed by the 1D NH Bloch bands, but also precisely captured the transition states between them. Our results may open a pathway for exploring and exploiting a wide range of NH topological effects in quantum and classical systems, such as the NH band theory that is still in its infancy.

*Tight-binding model for arbitrary Bloch braids.* Several theoretical models have been proposed for revealing the braiding and knot topology of NH Bloch bands [12,20,25,49,50]. Here we modify the simple model raised by Hu et al. [25] into a form more friendly to acoustic implementation (see *Supplemental Material* [51] for details). As sketched in Fig. 2(a) (top), the orbitals 1 and 2 (of zero onsite energy) are coupled by a reciprocal intracell hopping $t_0$ plus a unidirectional intercell hopping $t_m$ that spans $m$ lattices. For simplicity, here we assume $t_0 > 0$, $t_m \in \mathbb{R}$, and $m > 0$. The Hamiltonian of this lattice model reads

$$\mathbf{H}_{(m)} = \begin{pmatrix} 0 & t_0 \\ t_0 + t_m e^{imk} & 0 \end{pmatrix}, \quad (1)$$

which gives complex-valued energy spectra $E_\pm(k) = \pm\sqrt{t_0(t_0 + t_m e^{im})}$ and associated right eigenvectors $|\psi_\pm(k)\rangle = \left(1, \pm\sqrt{1 + (t_m/t_0)e^{imk}}\right)^T$. For any two separable energy bands, i.e., $E_+(k) \neq E_-(k)$ for all momenta $k \in [0, 2\pi]$, a topological invariant $v$ ($v \in \mathbb{Z}$) can be defined to classify their braiding topology,

$$v := \frac{1}{2\pi i} \oint_0^{2\pi} \left(\frac{d\ln E_+}{dk} + \frac{d\ln E_-}{dk}\right) dk, \quad (2)$$

which is isomorphic to the braid group $\mathbb{B}_2$ and essentially describe the braiding degree of the two complex bands [54]. Intuitively, it is equivalent to the total winding number of the two bands around the origin of the complex $E$-plane. Note that nontrivial braid topology occurs only in the NH systems, since the braiding degree $v = 0$ is preserved for all gapped Hermitian systems.

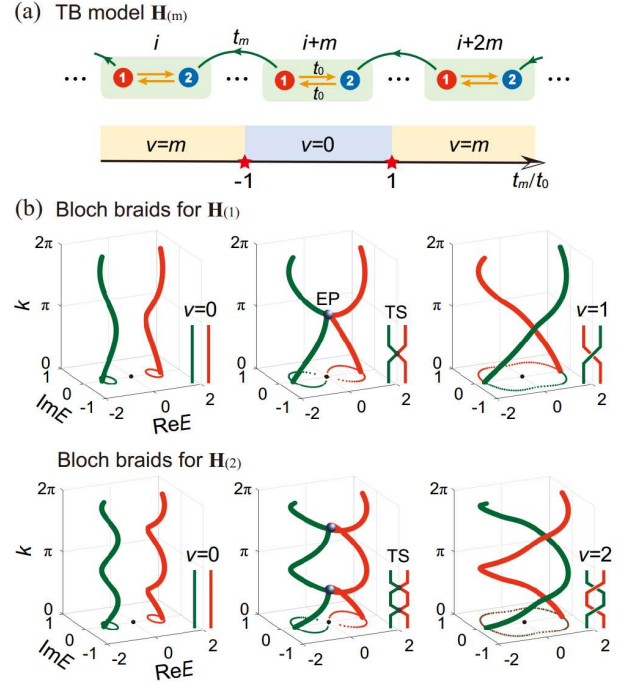

FIG. 2. 1D NH lattice model and Bloch braids. (a) Top: a sketch of the two-band model with reciprocal intracell hopping $t_0$ and nonreciprocal intercell hopping $t_m$. Bottom: the phase diagram exhibiting two topologically distinct braiding phases $v = 0$ and $v = m$, as well as their transitions (stars). (b) Representative complex band structures for the braiding models $H_{(1)}$ and $H_{(2)}$, each accompanied by an abstracted braid diagram. The data are calculated with fixed $t_0 = 1.0$ and varied $t_{1,2} = 0.5$ (left), 1.0 (middle), and 1.5 (right). The thin dots on the complex $E$-plane indicate the projected Bloch bands. The black spheres highlight the EPs emerged in transition states (TS).

Figure 2(a) (bottom) shows the phase diagram for our lattice model $\mathbf{H}_{(m)}$. It hosts two topologically distinct braiding phases $v = 0$ and $v = m$, associated to the parametric regions $|t_m/t_0| < 1$ and $|t_m/t_0| > 1$, respectively. At the phase boundary $t_m/t_0 = \pm 1$, the transition state features $m$ EPs at the isolated momenta satisfying $e^{imk} = \mp 1$, which results in a failure of the definition for the topological invariant $v$. Note that the Bloch braid of opposite handedness ($v = -m$) can be realized by simply reversing the coupling direction of $t_m$. Therefore, our model can not only realize an arbitrary nontrivial element of $\mathbb{B}_2$, but also its transition to the trivial phase ($v = 0$). See more discussions in *Supplemental Material* [51]. To illustrate the above picture, in Fig. 2(b) we demonstrate the complex band structures for two representative cases, $m = 1$ and $m = 2$. By simply inspecting the total winding numbers of the projected energy bands around $E = 0$, one may identify the trivial (left, $v = 0$) and nontrivial (right, $v = m$) braiding phases. These band-separable phases are bridged by the transition state (middle) accompanying a single EP at $k = \pi$, or two isolated EPs at $k = \pi/2$ and $3\pi/2$. Although not the focus of this work, we notice that all the phases will exhibit NH skin effect in a boundary-truncated lattice because of the presence of point gaps in their projected band structures onto the complex $E$-plane [25,44,45].

*Acoustic realization of the topological Bloch braids.* Experimentally, we employ a simple binary cavity-tube structure [28,55] (equipped with a unidirectional coupler) to realize the 1D NH model in synthetic space. Figure 3(a) shows our experimental setup. The acoustic sample consists of two identical air cavities, which have a fundamental dipole resonance frequency $\omega_0$ and an intrinsic loss $\gamma_0$. They are connected by two narrow tubes to produce the reciprocal intracell coupling $t_0$. A unidirectional coupler is elaborately designed for achieving a nonreciprocal hopping $\tilde{\kappa} = \rho e^{i\theta}$ with tunable amplitude $\rho$ and phase $\theta$. (Markedly different from the previous experimental implementation with fixed $\theta$ [28], here the capability of simultaneously tuning $\rho$ and $\theta$ plays a key role for unveiling the NH band topology.) The workflow can be introduced as follows: the sound signal in the cavity 1 is picked up by a microphone (labeled with $D_{UC}$), modulated by an amplifier and a phase shifter, and output into the cavity 2 through a loudspeaker ($S_{UC}$). To experimentally retrieve the intrinsic parameters of our acoustic system and realize the controllable nonreciprocal coupling, we measure the transmission responses $S_{21}(\omega)$ and $S_{12}(\omega)$ without and with turning on the external unidirectional coupler, where the subscripts $i$ and $j$ in $S_{ij}(\omega)$ denote the cavities inserted with the acoustic detector $D$ and source $S$, respectively.

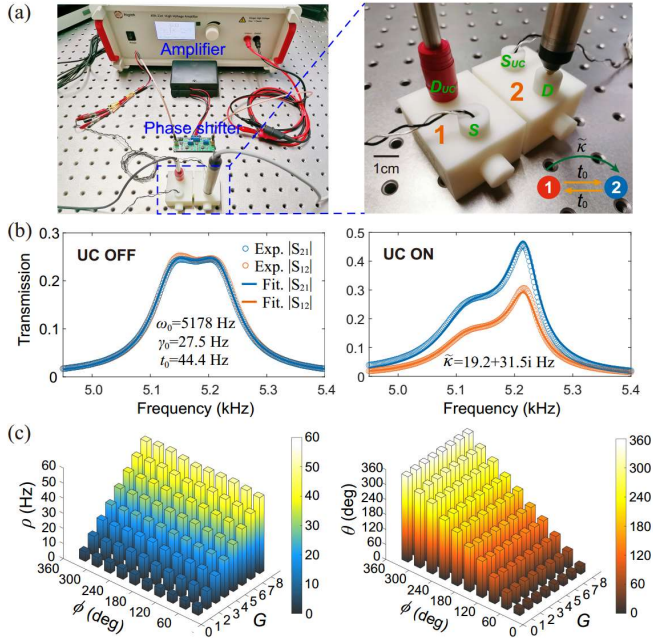

FIG. 3. Acoustic realization of the unidirectional complex coupling $\tilde{\kappa}$. (a) Experimental setup. The acoustic cavities 1 and 2 emulate the orbitals and the narrow tubes in between (invisible here) mimic the reciprocal coupling $t_0$. A controllable unidirectional coupler (UC), consisting of a microphone $D_{UC}$, an amplifier, a phase shifter, and a loudspeaker $S_{UC}$, is used to generate the complex coupling $\tilde{\kappa} = \rho e^{i\theta}$. A source $S$ and a detector $D$ are used to excite and detect acoustic transmission responses, respectively. (b) Measured and fitted transmission spectra for the cases of turning off ($\tilde{\kappa} = 0$) and on ($\tilde{\kappa} \neq 0$) the unidirectional coupler. (c) Linear relations identified for $\rho(G)$ and $\theta(\phi)$, where $G$ and $\phi$ represent the gain factor and the phase shift controlled by the unidirectional coupler.

The left panel of Fig. 3(b) shows the measured transmission responses with the unidirectional coupler turned off. In this case, the system remains reciprocal ($\tilde{\kappa} = 0$) and gives nearly overlapped $|S_{21}(\omega)|$ and $|S_{12}(\omega)|$. By fitting the spectra with the established coupled-mode theory [51], we obtain the intrinsic parameters $\omega_0 \simeq 5178$ Hz, $\gamma_0 \simeq 27.5$ Hz, and $t_0 \simeq 44.4$ Hz under reciprocity. Turning on the unidirectional coupler, we exemplify the transmission spectra under the gain factor $G = 5.5$ and phase shift $\phi = 60°$ [Fig. 3(b), right]. As a manifestation of non-reciprocity, the transmission responses $|S_{21}(\omega)|$ and $|S_{12}(\omega)|$ become visibly different and even highly asymmetric in their line shapes. With the already accessed intrinsic parameters, the nonreciprocal coupling can be fitted as $\tilde{\kappa} \simeq (19.2 + 31.5i)$ Hz. By repeating this procedure with different $G$ and $\phi$, we realize a wide range of nonreciprocal couplings [Fig. 3(c)]. Interestingly, the results demonstrate nearly independent linear relations $\rho(G) \simeq 6.44G$ Hz and $\theta(\phi) \simeq \phi − 0.92°$, associated with fitting errors of 3.1% and 2.1%, respectively. This enables an elegant realization of the lattice model $\mathbf{H}_{(m)}$ by synthesizing the momentum-resolved nonreciprocal coupling $t_m e^{imk}$ with $\tilde{\kappa}(G, \phi)$ (see *Supplemental Material* [51]).

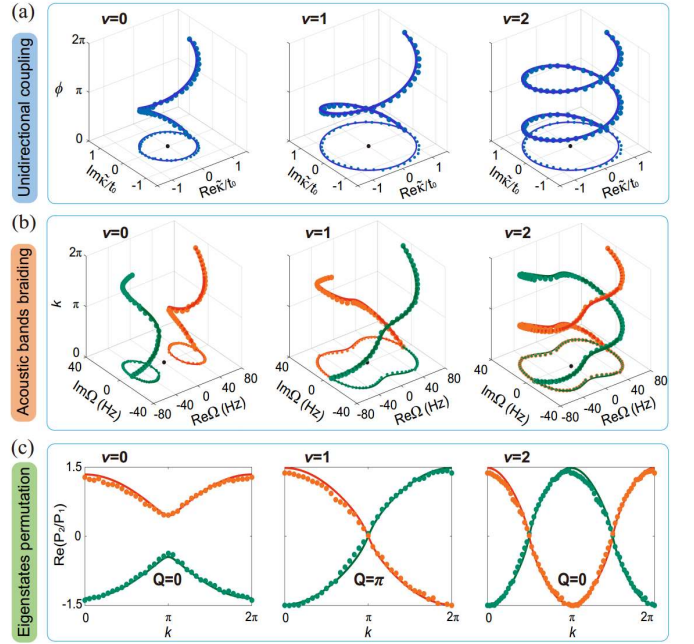

FIG. 4. Acoustic realization of the topological Bloch braids. (a) Unidirectional coupling $\tilde{\kappa}$ designed for achieving the different Bloch braids in (b). (c) Permutation parities exhibited in the $k$-evolved eigenstates. All experimental data (dots) match well the theoretical predictions (lines).

Now we experimentally demonstrate the topological Bloch braids for $\nu = 0,1,2$ in synthetic space, where the momentum $k$ is synthesized by the mappings $\rho \to t_m$ and $\theta \to mk$, together with the independent linear relations $\rho(G)$ and $\theta(\phi)$. First, we fix the gain factor $G = 5.5$ and tune the phase shift $\phi$ from $0°$ to $360°$ at a step of $10°$. The fitted $\tilde{\kappa}$ [Fig. 4(a), left] emulates the term $t_1 e^{ik}$ with $t_1 \simeq 0.8 t_0$ in $\mathbf{H}_{(1)}$. Substituting $\tilde{\kappa}$ into the Hamiltonian, we obtained two complex bands through solving the eigenvalues. As expected, the two separable NH bands [Fig. 4(b), left] do not braid around each other, exhibiting a trivial braiding topology of $\nu = 0$. (For clarity, the complex band frequency $\Omega$ is measured from the acoustic 'onsite energy' $\omega_0 - i\gamma_0$.) By contrast, turning $G$ up to 8.5 while preserving the variation of $\phi$ yields a nontrivial braid of $\nu = 1$ [Fig. 4(b), middle], since the increased $\rho$ gives effectively $t_1 \simeq 1.2 t_0$ in $\mathbf{H}_{(1)}$ [Fig. 4(a), middle]. Finally, to realize the acoustic braid of $\nu = 2$, we maintain $G = 8.5$ but vary $\phi$ from $0°$ to $720°$. The resultant unidirectional coupling $\tilde{\kappa}$ [Fig. 4(a), right] gives $t_2 \simeq 1.2 t_0$ in $\mathbf{H}_{(2)}$, and thus the two bands braid twice around each other [Fig. 4(b), right]. Given the periodicity of the 1D Brillouion zone, the above Bloch braids form an unlink, an unknot, and a Hopf link in turn. The braiding topology of complex bands can also be demonstrated through eigenfunctions. It of particular interest that an extra $\mathbb{Z}_2$ knot invariant, $Q$, can be defined with a global biorthogonal Berry phase [25]. Physically, the topological invariant $Q = 0$ ($\pi$) characterizes the even (odd) permutation parity of the eigenstates $|\psi_{\pm}\rangle$ during the bands braiding from $k = 0$ to $2\pi$. (Note that all gapped Hermitian systems are topologically trivial in this classification.) Visually, the knot invariants 0 and $\pi$ distinguish the braiding configurations of links and knots, respectively. To experimentally evidence this exotic $\mathbb{Z}_2$ knot topology, we detect the transmission responses $P_1 = S_{12}$ and $P_2 = S_{22}$ at real eigenfrequency, and extract the real-part information of the eigenstates, $\mathrm{Re}(P_2/P_1)$ [51]. Figure 4(c) shows the $k$-evolution of the eigenstates for the above three systems. It is easy to find that the eigenstates at $k = 0$ and $2\pi$ are interchanged for the case of $\nu = 1$ (unknot), as a reflection of the nontrivial knot invariant $Q = \pi$. By contrast, no permutation of the eigenstates emerges in the cases of $\nu = 0$ (unlink) and $\nu = 2$ (Hopf link), manifesting a trivial knot invariant $Q = 0$.

Without details provided here, we have also identified the transition state between the Bloch braids of $\nu = 0$ and $\nu = m$, through simply obliging $\tilde{\kappa}(k) = t_m e^{imk}$ with $t_m = t_0$ (see Supplemental Material [51]). Unambiguously, our experimental results demonstrate the simultaneous coalescence of the eigenvalues and eigenstates at the predicted EP momenta. Note that the intriguing transition state between the Bloch braids, which goes beyond the description of the braid group $\mathbb{B}_2$, has not been unveiled experimentally elsewhere.

*Topological transition between two arbitrary Bloch braids*. So far we have demonstrated the capability of the lattice model $\mathbf{H}_{(m)}$ in realizing the band-separable braids with $\nu = 0$ and $\nu = m$, as well as the band-inseparable intermediate state between them. As illustrated in Fig. 5(a), below we consider a lattice model with one more unidirectional coupling that spans $n$ lattices, $t_n$ ($n > m$). The Hamiltonian reads

$$\mathbf{H}_{(m,n)} = \begin{pmatrix} 0 & t_0 \\ t_0 + t_m e^{imk} + t_n e^{ink} & 0 \end{pmatrix}. \quad (3)$$

Comparing with $\mathbf{H}_{(m)}$, the introduction of the extra long-range coupling will enrich the phase diagram beyond doubt. Particularly, this extension enables to realize a phase transition between two arbitrary Bloch braids $\nu = m$ and $n$ [51], where the long-range couplings demanded in $\mathbf{H}_{(m,n)}$ can be readily incorporated into our acoustic experiments with the concept of synthetic dimensions. As an example, here we consider $m = 1$ and $n = 3$. The phase diagram [Fig. 5(b)] shows four distinct topological phases (with $\nu = 0 \sim 3$), where the phase boundaries can be derived analytically [51]. Below we focus on the direct phase transition highlighted by the path $A \to B \to C$, where the state $B$ corresponds to a transition state between the nontrivial braiding phases $\nu = 1$ and $\nu = 3$.

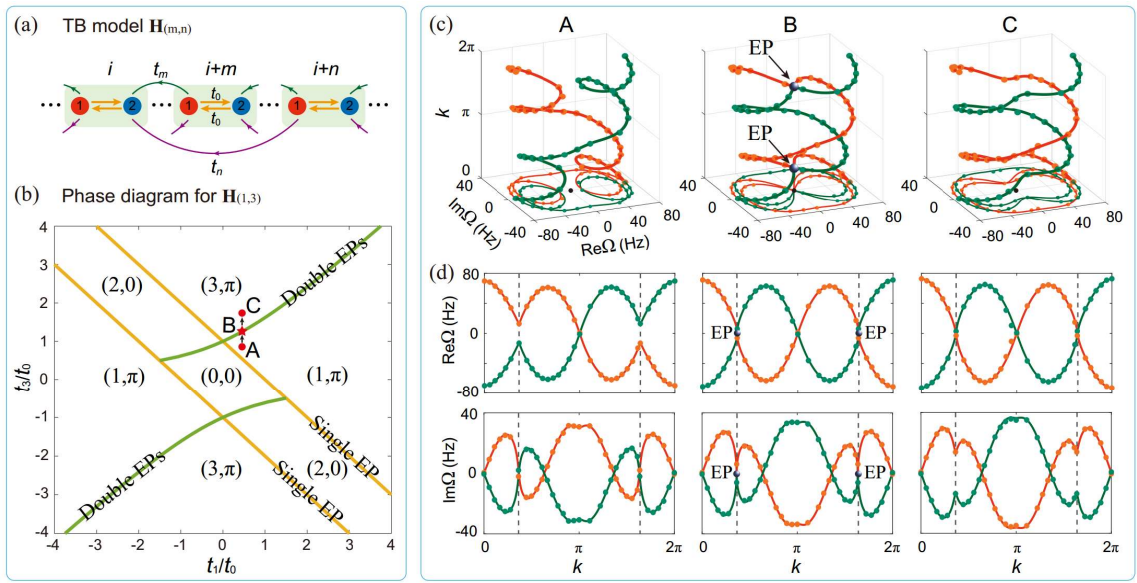

FIG. 5. Topological transition between two arbitrary braiding phases. (a) Tight-binding model $\mathbf{H}_{(m,n)}$. (b) Phase diagram exemplified for $\mathbf{H}_{(1,3)}$.

(c) Experimentally measured complex band structures (dots) for the states *A*, *B*, and *C*, compared with their theoretical predictions (lines). (d) Same as (c), but plotted with the real (top) and imaginary (bottom) frequencies separately. The dashed lines highlight the momenta where the EPs emerge.

To experimentally observe the phase transition, we tune the gain factor $G$ and phase shift $\phi$ simultaneously to ensure $\tilde{\kappa}(k) = t_1 e^{ik} + t_3 e^{i3k}$ in $\mathbf{H}_{(1,3)}$, where $k$ evolves from $0$ to $2\pi$. Specifically, the states *A*, *B*, and *C* share the same $t_1 = 0.37 t_0$ but have distinct $t_3$, i.e., $1.1 t_0$, $1.2 t_0$, and $1.3 t_0$, respectively. Figure 5(c) presents the measured complex band structures for the three states, following the experimentally fitted complex couplings $\tilde{\kappa}(k)$ like Fig. 4(a) (see Supplemental Material [51]). As expected, the double bands of the state *A* (*C*) are separable and braided each other once (three times), exhibiting a feature of the unknot (Trefoil knot). By contrast, for the transition state *B*, the double bands tend to touch at the predicted EPs with $k \simeq 0.45\pi$ and $1.55\pi$. The above transition process can be further unveiled from the individual plots with real and imaginary frequencies [Fig. 5(d)]. Comparing with those of the states *A* (left) and *C* (right), the real and imaginary bands of the transition state *B* (middle) contact at the two EP momenta simultaneously. Thus far, our experiments, reproducing precisely the theoretical predictions, unambiguously demonstrate a transition process from an unknot ($\nu = 1$) to a Trefoil knot ($\nu = 3$).

*Conclusion.* Resorting to the concept of synthetic dimensions, we have experimentally demonstrated some representative acoustic NH Bloch braids, as well as the highly-desired topological transition states in between. The former with separable complex bands exhibits a $\mathbb{B}_2$ braid group topology dictated by mathematical knot theory, while the latter, featuring exceptional band degeneracies in complex *E-k* space, exceeds the description of the braid group. As a building block, our two-band models and acoustic experiments can be easily extended to construct multi-band braiding structures and probe the intriguing non-Abelian effects tied to the braid group $\mathbb{B}_N$ ($N > 2$) [50,54]. More challengingly, our study can be generalized to higher-dimensional and symmetric systems, where the intricate interplay of the band braiding, eigenstate topology, and symmetry gives rise to richer but unexplored topological phenomena [29,30]. Given extra time modulation of the Hamiltonian parameter, it is also highly interesting to explore the dynamic evolution of braiding and knots, which involves mix, superposition, and interaction of different *k*-modes [25,56]. The last but not least, our experimental technique for achieving fully controllable complex couplings enables to design various NH acoustic metamaterials with unprecedented properties and functionalities [57-63].

### ACKNOWLEDGEMENTS

This work is supported by the National Natural Science Foundation of China (Grant No. 11890701, 12104346, 12004287), and the Young Top-Notch Talent for Ten Thousand Talent Program (2019-2022).